# Transport and magnetic behavior under pressure and high-resolution photoemission studies of $Ce_2Rh_{0.7}Co_{0.3}Si_3$, an alloy on the verge of quantum critical point

K. Mukherjee, Kartik K Iyer, Swapnil Patil, K, Maiti, and E.V. Sampathkumaran
Tata Institute of Fundamental Research, Homi Bhabha Road, Mumbai-400005, India.

E-mail: kmukherjee@tifr.res.in
E-mail: sampath@tifr.res.in

**Abstract.** We report the influence of external pressure on the temperature dependence of magnetization and electrical resistivity as well as high-resolution photoemission studies for an alloy, $Ce_2Rh_{0.7}Co_{0.3}Si_3$, ordering magnetically below 3 K. It is found that the external pressure has the same effect as that induced by (further) Co substitution for Rh in the series, $Ce_2Rh_{1-x}Co_xSi_3$, resulting in qualitative changes in the features in the magnetic and transport data, with a suppression of magnetic ordering followed by quantum critical point effect. The high-resolution photoemission spectra reveal signature of Kondo feature even at ambient pressure. These findings support the validity of spin-density-wave picture in this series.

## 1. Introduction

There is a considerable debate in the current literature [1-3] on how Fermi surface evolves as one gradually transforms a magnetically ordered Ce/U intermetallic to non-magnetism. That is, there are some materials where magnetically ordered state before 'quantum critical point' (QCP) can be viewed as a spin-density-wave (SDW) state due to itinerancy of f-electrons (associated with the Kondo effect), while some other experiments support 'localized picture'. Recently, we addressed [4] this question in the solid solution, $Ce_2Rh_{1-x}Co_xSi_3$, forming in an $AlB_2$-derived hexagonal structure. $Ce_2RhSi_3$ [5] and $Ce_2CoSi_3$ [6] have been known to be antiferromagnetic (near $T_N=$) 7 K and non-magnetic respectively with quantum critical point (QCP) near $x=$ 0.6. The electrical resistivity ($\rho$) data showed distinct qualitative changes below $T_N$ with increasing Co content. In particular, apart from other features, there is a distinct upturn appearing in $\rho(T)$ below $T_N$ for $x=$ 0.3 which disappeared with increasing $x$ near QCP (see figure 1). This was taken as an evidence for a gradual transformation of Fermi surface as QCP is approached and hence for "spin-density-wave picture" (rather than "local moment picture") for this system. In order to support this line of argument, we considered it worthwhile to probe how external pressure influences the transport behavior for such a composition. Therefore, we carried out high pressure (< 15 kbar) *ac* resistivity ($\rho$) and *dc* magnetization (*M*) studies on the alloy, $Ce_2Rh_{0.7}Co_{0.3}Si_3$. In addition, we have performed high-resolution (< 5 meV) photoemission studies as a function of temperature ($T$) for this composition to see the evolution of the Kondo features. The observed findings support the views based on spin density wave scenario as discussed earlier for this family [4].

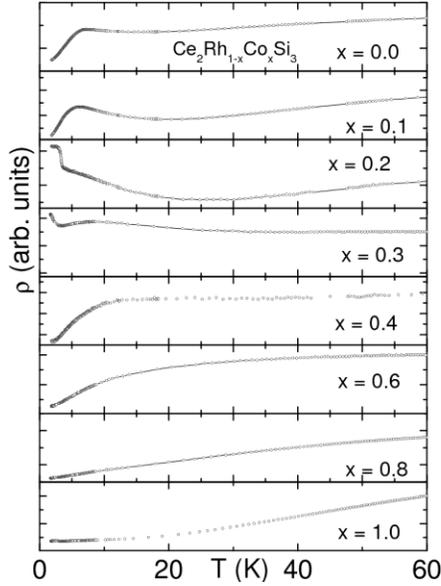

**Figure 1.** Electrical resistivity behavior (linear scale) at low temperatures for the alloys, $Ce_2Rh_{1-x}Co_xSi_3$. The data were taken from Ref. 4.

## 2. Experimental details

The polycrystalline specimen employed in the present investigation is the same as that in our previous studies. The $M$ studies under pressure (≤10 kbar) were carried out (1.8-300 K) in a magnetic field of 5 kOe with a SQUID magnetometer (Quantum Design) using a pressure cell procured from EasyLab Technologies Ltd (UK) in a hydrostatic pressure medium of daphne oil. The $\rho(T)$ data in zero field and in field (<140 kOe) were carried out (1.8-300 K) with a physical property measurements system (Quantum Design) in a hydrostatic medium of a mixture of pentane and iso-pentane up to ≤15 kbar. The electrical contacts of the leads to the specimen were made by a conducting silver paste. Photoemission measurements were carried out on scraped sample surface using an electron analyzer SES2002 and monochromatic photon sources at base pressure of $3 \times 10^{-11}$ torr and an energy resolution set to 5 meV.

## 3. Results and discussion

The results of magnetization measurements measured in a field of 5 kOe at low temperatures (20 K) are shown in figure 2*a* in the form of a plot of *M/H versus T* to enable us to draw qualitative conclusions on the magnetic ordering behavior. The feature due to magnetic ordering appearing as a change in slope at 3 K for ambient pressure conditions persists at 5 kbar as well, though it is prominently seen in the form of a peak for this pressure. For a slight increase of pressure to 6 kbar, interestingly, this maximum disappears. For an application of 10 kbar, it appears that the Curie Weiss variation persists down to 1.8 K, as though the magnetic ordering is suppressed. In figure 2(*b*) and 2(*c*), we show the $\rho(T)$ behavior. It is clear that the upturn below 3 K (due to possible

SDW gap) seen at ambient pressure vanishes at 5 kbar resulting in a positive d$\rho$/d$T$ at very low temperatures, as though there is a change in the Fermi level position with respect to the gap. However, the feature (a drop) near 8 K attributable to the Kondo coherence effect is enhanced marginally by a few degrees, which appears to smear out the double-peaked structure (characterizing Kondo lattices) in $\rho(T)$. This smearing is very significant for higher pressures (10 and 15 kbar) resulting in a single broad peak (around 60 and 70 K respectively). It is well-known that an increase of this (Kondo coherence) peak with temperature is the characteristic influence of pressure in the Ce-based Kondo lattices. A noteworthy feature is that $\rho$ appears to vary linearly below about 8 K for $P$ = 10 kbar (the pressure at which magnetic ordering appears to get suppressed), as though at this pressure-induced QCP, non-Fermi-liquid behavior develops. For the higher pressure (15 kbar), quasi-linearity with a higher exponent (~ $T^{1.2}$) is distinctly observed. All these features are qualitatively the same as that seen for gradual replacement of Rh by Co, thereby endorsing the arguments in Ref. 4. An additional feature we stress is that the quasi-linear behavior seen in $\rho(T)$ at 15 kbar interestingly becomes linear if $\rho$ is measured in the presence of a large field (say, e.g., 130 kOe) as shown in figure 2(*c*), as though there is a tendency to move away from Fermi-liquid behavior at very high fields for the pressurized specimen.

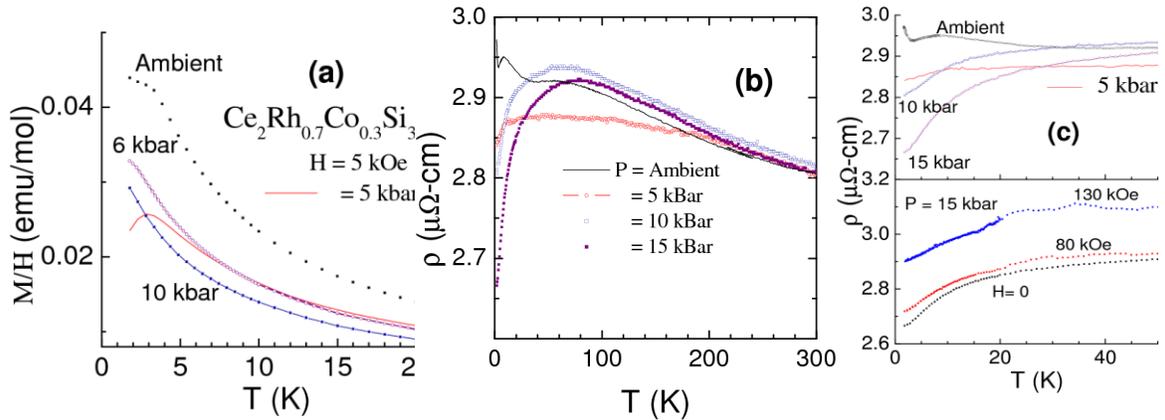

**Figure 2.** (*a*) Temperature dependence of magnetization divided by magnetic field, under external pressure of the alloy, Ce$_2$Rh$_{0.7}$Co$_{0.3}$Si$_3$. (*b*) Electrical resistivity as a function of temperature for the same specimen under pressures is plotted. (*c*) The $\rho$ data under pressure as well as in magnetic field at low temperatures are plotted.

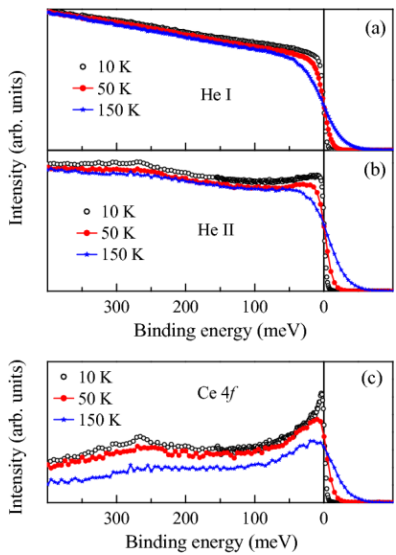

9.

**Figure 3.** (*a*) He I and (b) He II spectra of $Ce_2Rh_{0.7}Co_{0.3}Si_3$ at different temperatures. (c) (He II - He I) spectra presenting the Ce 4*f* signal at different temperatures.

In figure 3, we show the high resolution He I and He II spectra collected at different temperatures. While the He I spectra exhibit a typical temperature induced change at the Fermi level following closely the Fermi-Dirac distribution function, the He II spectra reveal two distinct features, one at the Fermi level and the other around 280 meV binding energy. The relative photoemission cross section of Ce 4*f* states is significantly larger in He II spectra compared to that in He I spectra. Thus, the features in the He II spectra can be attributed to the signature of Ce 4*f* states (the feature at 280 meV is spin-orbit split Ce 4*f* signal) that can be delineated by subtracting He I spectra from the He II spectra. The subtracted spectra shown in figure 3(*c*) exhibit distinct signatures of the growth of Ce 4*f* features characteristic of Kondo resonance. The observation of this feature at ambient pressure for an alloy which is antiferromagnetic is interesting and demonstrates the persistence of Kondo compensation effect within the magnetically ordered region in Doniach's magnetic phase diagram, thereby supporting SDW scenario.

**4. Conclusion**
We have reported how external pressure influences magnetic and transport behavior of a magnetically ordering alloy, *x*= 0.3. In addition, high-resolution photoemission spectra down to low temperatures were also measured which show the signature of Kondo resonance feature at ambient pressure. These findings clearly support the validity of SDW picture in this family of materials.